\documentclass[11pt]{amsart}

\usepackage{amssymb}
\usepackage{amsmath,amsthm,amsfonts}  
 \usepackage{graphicx}
\usepackage{amssymb}
\usepackage{amsmath}
\usepackage{amsthm}
\usepackage{pdfsync}
\usepackage{fancyhdr}
\usepackage[applemac]{inputenc}

\usepackage[default]{frcursive}
\usepackage[T1]{fontenc}

\usepackage{color}
\usepackage{xcolor}
\usepackage[breaklinks,colorlinks,backref]{hyperref}
\hypersetup{
    colorlinks, 
    linktoc=all, 
    linkcolor=red,  
  citecolor=hyptxt,
  urlcolor=blue}
  \hypersetup{
  citebordercolor=,
  filebordercolor=green,
  linkbordercolor=blue
}
\definecolor{hyptxt}{rgb}{0.7, 0.4, 0.9}
\usepackage{bibtopic}

\usepackage[full]{textcomp} 
     \usepackage{fbb} 
     \usepackage[scaled=.95]{cabin}
     \usepackage[varqu,varl]{inconsolata}
     \usepackage[libertine,bigdelims,vvarbb]{newtxmath}
     \usepackage[cal=boondoxo]{mathalfa}
     \usepackage[T1]{fontenc}
\useosf

\definecolor{hervecolor}{rgb}{0.8,0,0.7}
     
\newcommand{\ket}[1]{|\kern.3ex#1\kern.3ex\rangle}
\newcommand{\bra}[1]{\langle\kern.3ex #1 \kern.3ex|}
\newcommand{\scalar}[2]{\langle\kern.3ex #1 \kern.3ex|\kern.3ex#2\kern.3ex\rangle}

\newcommand{\ii}{\mathsf{i}}

\def\R{\mathbb{R}}

\def\lg{\langle }
\def\rg{\rangle }

\def\ud{\mathrm{d}}
\def\sfH{\mathsf{H}}

\def\sfP{\mathsf{P}}

\def\mfF{\mathfrak{F}}

\def\mFs{\mathfrak{F_s}}

\def\vr{\pmb{r}}
\def\va{\pmb{a}}
\def\vb{\pmb{b}}
\def\vrp{\pmb{r}^{\prime}}
\def\vrpp{\pmb{r}^{\prime\prime}}

\def\vz{\pmb{0}}

\numberwithin{equation}{section}

\begin{document}
\date{\today}
 
\title[Weyl-Heisenberg Integral quantization]{From classical to  quantum models: the regularising  r\^ole of integrals, symmetry and probabilities}
\author[J.-P. Gazeau]{
Jean-Pierre Gazeau$^{\mathrm{a,b}}$}

\address{\emph{  $^{\mathrm{a}}$ APC, UMR 7164,}\\
\emph{Univ Paris  Diderot, Sorbonne Paris Cit\'e}  
\emph{75205 Paris, France}} 

\address{\emph{$^{\mathrm{b}}$ Centro Brasileiro de Pesquisas F\'{\i}sicas } \\
\emph{Rua Xavier Sigaud 150, 22290-180 - Rio de Janeiro, RJ, Brazil  }}

\email{e-mail:
gazeau@apc.in2p3.fr}

{\abstract{In physics, one is often misled in thinking that the mathematical model of a system is part of or is  that system itself. Think of expressions commonly used in physics  like ``point'' particle, motion ``on the line'', ``smooth'' observables, wave function, and even ``going to infinity'', without forgetting perplexing phrases like ``classical world'' versus ``quantum world''.... On the other hand, when a mathematical model becomes really inoperative  in regard with correct predictions, one is forced to replace it with a new one. It is precisely what happened with the emergence of quantum physics. Classical models were (progressively) superseded by quantum ones through quantization  prescriptions.  These procedures appear often as ad hoc recipes. In the present paper, well defined quantizations,  based on integral calculus and  Weyl-Heisenberg symmetry, are described in simple terms through one of the most basic examples of mechanics. Starting from  (quasi-) probability distribution(s) on the Euclidean plane viewed as the phase space for the motion of a point particle on the line, i.e., its classical model, we will show how to build corresponding quantum model(s) and associated probabilities (e.g. Husimi) or quasi-probabilities (e.g. Wigner) distributions. We highlight  the regularizing r\^ole of such procedures  with  the familiar example of the  motion of a particle with a variable mass and submitted to a step potential.  
 }}

\maketitle

\tableofcontents

\section{Introduction}
\label{intro}


\subsubsection*{A world of mathematical models for one ``thing'' in the ``World''}

The physical laws are expressed in terms of combinations of mathematical symbols,  numbers, functions, geometries, relations ...
These combinations take place  within a mathematical model for the system, a part of the so-called objective reality under consideration.  
Such a  language and related concepts are in constant development since the set of phenomenons which are accessible to our scientific understanding is constantly broadening, or at least, reshaped.  Now, a model for a system is usually scale dependent. It depends on a ratio of physical, i.e., measurable, quantities, like amounts of substance, lengths, time(s), sizes, impulsions, actions, energies ... In certain cases, a radical change of scale, radical in the mathematical sense of limit, for a model amounts to ``quantize'' or ``de-quantize''. As a matter of fact, one decides on the validity of a classical model versus a quantum one for a given physical system if  the action(s) which is (are) characteristics of the latter, e.g. $\sqrt[3]{\mbox{\textit{spatial size}}}\times$\textit{momentum} or \textit{energy}$\times$\textit{time duration} or \textit{angular momentum},  is (are) $\gg\hbar$. One changes  perspective, one can say that our understanding changes its  (mathematical) glasses!

\subsubsection*{Exactness of  models and probabilities}

Nothing is mathematically exact from the physical point of view.  A mathematical model in Physics is never for ever. Its suitability is time dependent because it is scale dependent. In order to be adjusted to experimental  observations and predictions, it has to be modified more or less radically, even radically changed. In the relation \textit{modeler} $\leftrightarrow$ \textit{modelled object}, there is probability, $\sim$  degree of epistemic confidence in the suitability of the model. Hence emerges the necessity of some \textit{coarse-graining} of the initial mathematical model supposed to describe  a certain ontic entity or fact. 

For instance,  irrational numbers are far beyond human perception but physical laws are usually (since  a few centuries) expressed in terms of real numbers, $\R$, built from limit notions (limit of Cauchy sequences of rational numbers). Now,  an infinite amount of energy would be needed to measure the location of one point on the real line! A coarse-graining, or \textit{quantization} in the sense of signal analysis, is naturally requested on an operational level.

Even the notion of contextuality, which describes how or whether the details of an observation affect what is observed, cannot be dissociated from the mathematical model used for the definition of the object, its observation, and its interpretation. 

\begin{figure}[htb!]
\begin{center}
\setlength{\unitlength}{0.1cm} 
\begin{picture}(60,60)
\put(0,30){\vector(1,0){60}} 
\put(30,10){\vector(0,1){45}}
\put(30,30){\vector(1,1){15}} 
\put(27, 27){\makebox(0,0){$O$}} 
\put(60, 27){\makebox(0,0){$q$}}
\put(27, 55){\makebox(0,0){$p$}} 
\put(51, 41){\makebox(0,0){$\vr:=(q,p)
$}} 
\put(45, 45){\makebox(0,0){$\bullet$}} 
\end{picture}
\caption{Set of initial conditions $\sim$ phase space for the motion of a point particle on the real line.}
\label{phasespace}
\end{center}
\end{figure}
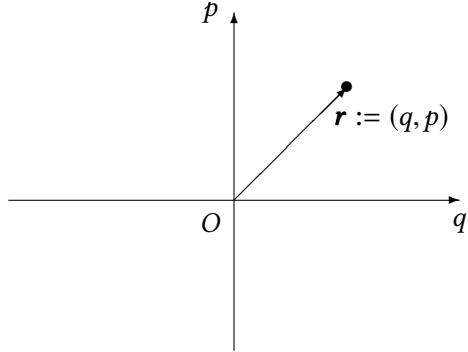

The aim of this paper is to  show  that the construction of a quantum model from a classical one pertains, in a certain sense, to that type of coarse-graining procedure. The procedure is illustrated with one of the most elementary examples in mechanics, namely the motion of a \textit{point} particle on the \textit{straight} $\sim$ \textit{real} line $\sim$ $\R$ , for which the  phase space is the plane shown in Figure \ref{phasespace}.
\begin{equation}
\label{phaspace1}
\R^2= \{ \vr = (q,p)\, , \, q,p \in \R\}  \, .  
\end{equation}
 We then establish the quantum versions of this classical model by developing an approach combining  probability (the coarse-graining)   with symmetry and integral calculus. 
 
A  part of this work will  certainly appear familiar to most of the readership. However we have chosen to present the material, e.g., Weyl-Heisenberg and Galilean symmetries, basic rules of quantum formalism, in a somewhat uncommon and  self-contained way, and we want to emphasize the benchmark role played by obvious symmetry requirement(s) in  any quantization  procedure. 

In Section \ref{stquant} are recalled the essential features of the quantum formalism and the way it is established from Hamiltonian classical mechanics. A  survey of various quantization methods is also sketched. 
The covariant Weyl-Heisenberg integral quantization is the subject of Section \ref{CIQWH}. Starting from the translation symmetry of the Euclidean plane, we show how to reach its non commutative  representation underlying the corresponding quantum model of the phase space through integral maps involving operator-valued measures. 
In Section \ref{permissu} the above procedure is implemented in the case of elementary functions on the phase space, namely coordinates, quadratic expressions, functions of $q$ (resp. $p$) only. We point out the similarities and differences between their respective quantum versions in function of the operator-valued measure underlying a specific integral quantization.   We show in Section \ref{mvarhamil} that the method easily applies to  Hamiltonian expressions constrained by  the so-called shadow Galilean invariance, an important notion that we explain in detail because it fully justifies the concept of variable mass.  
We give in  Section \ref{weight} a few examples of these operator-valued measures, and discuss about their relevance in dealing  with  specific classical functions. 
Section \ref{probcont} is devoted to the description of the probabilistic aspects of our quantization procedure and its  reversal under the form  of \textit{quantum phase space portraits}. This leads to an  interesting analogy with  the intensity of a  diffraction pattern resulting from the  coarse graining of the idealistic phase space $\R^2$. 
The general method is illustrated in Section \ref{exreg} with the textbook model of a variable mass particle whose  one-dimensional motion is constrained  by a  potential barrier.  
We conclude in Section \ref{conclu} by giving some insight about the generalisation of the approach to phase spaces presenting different  symmetries, and to manifolds embedded in higher dimensional phase spaces.

\section{Considerations  on standard  and other quantizations }
\label{stquant}

The basic, or so-called canonical, quantization procedure 
  starts from the phase space  $\R^2$,
\begin{align}
\R^2\ni (q,p)  &\mapsto \ \mbox{self-adjoint}\ (Q,P)\,, \quad [Q,P] = \ii \hbar I\, ,\\
f(q,p) &\mapsto f(Q,P) \mapsto (\mathrm{Sym}f)(Q,P)\, ,
\end{align}
where $\mathrm{Sym}$ stands for a certain choice of symmetrisation  of the operator-valued function. 
We remind that $[Q,P] = \ii \hbar I$ holds true with (essentially) self-adjoint $Q$, $P$,  only if both have continuous spectrum $(-\infty,+\infty)$. We also remind that a quantum observable is an essentially ($\sim$ no ambiguity) self-adjoint operator in the Hilbert space $\mathcal{H}$ of quantum states,  since the key for (sharp) quantum measurement is encapsulated in the spectral theorem for a bounded or unbounded self-adjoint operator $A$. The latter  asserts that $A$ has a real spectrum  with integral representation 
\begin{equation}
A= \int_{\Sigma\subset \R} \lambda\,\ud E_A(\lambda)\, , 
\end{equation}
involving  a normalised projective-operator measure  $\ud E_A(\lambda):= E_A((\lambda, \lambda + \ud \lambda))$. The expression \textit{projective} means that 
\begin{equation}
E_A(\Sigma \cap\Sigma^{\prime}) = E_A(\Sigma)\, E_A(\Sigma^{\prime})
\end{equation}
whereas normalisation means  resolution of the identity in $\mathcal{H}$,
\begin{equation}
\int_{\Sigma\subset \R} \ud E_A(\lambda) = I\, .  
\end{equation}
In this context, the position operator is self-adjoint  with spectrum $\R$
\begin{equation}
Q= \int_{\R} \lambda\, |\lambda\rg\lg\lambda|\, \ud\lambda\, ,  \qquad \lg x |\lambda\rg = \delta(x-\lambda)\, . 
\end{equation}
and the Hilbert space of quantum states is realized as functions $\psi(x)$, the wave functions,  which are square integrable on the \underline{spectrum} of the position operator $Q$. At the heart of the concept of localisability,  the variable $x$ has to be interpreted as a (measurable) element of the spectrum of $Q$: it is an essential part of the quantum model,  and not of the  classical one. Moreover,  its definition is not ambiguous in Galilean quantum mechanics \cite{inwig50,wightman62}. 
Accordingly, the action of $P$ on this Hilbert space results from $[Q,P] = \ii \hbar I$
\begin{equation}
P\psi(x) = -\ii \hbar\frac{\ud}{\ud x}\psi(x)\,. 
\end{equation}
This simple scheme presented in  many introductive textbooks to quantum mechanics immediately raises fundamental questions 
like what about singular $f$, e.g. the angle or phase $\arctan(p/q)$? What about other phase space geometries? Barriers or other  impassable boundaries? The motion on a circle (the question of quantum angle and localisation on the circle)? In a bounded interval? On the positive half-line (singularity at the origin)? ....
 Despite their elementary
aspects, these disturbing examples leave open many questions both on mathematical
and physical levels, irrespective of the manifold of quantization procedures, like
 Lagrangian \& Path Integral Quantization (Dirac 1932, Feynman, thesis, 1942).
 Or, after approaches by Weyl (1927), 
Groenewold (1946), Moyal
(1947),  
Geometric
Quantization,  Kirillov (1961)
Souriau (1966), Kostant (1970), 
 Deformation Quantization, Bayen, Flato, Fronsdal, Lichnerowicz, Sternheimer (1978),
Fedosov (1985), Kontsevich (2003),  
coherent state or anti-Wick or
Toeplitz quantization with Klauder (1961), Berezin (1974) 
and others..., see for instance   illuminating articles like \cite{berezin74}, comprehensive reviews like \cite{alienglis05} \cite{landsman06}, and  more recent volumes like \cite{combescure12} \cite{degosson16},  about these various approaches.  
Indeed, most of  these methods, despite their beautiful mathematical content, are often too demanding for some classical models to be consistently quantized.  

On the other hand, it is fair to acknowledge that canonical quantization is quasi-universally accepted in view of its numerous
experimental validations, one of the most famous and simplest one
going back to the early period of quantum mechanics with the quantitative
prediction (1925) of the isotopic effect in vibrational spectra of diatomic
molecules (see \cite{bergayou13} and references therein).  
These data validated the canonical quantization,
contrary to the Bohr-Sommerfeld ansatz (which predicts no isotopic
effect). 
Nevertheless this does not prove that another method of quantization
fails to yield the same prediction.  
Moreover  the canonical quantization
appears as  too rigid or even untractable in some circumstances, as was  underlined above. As a matter of fact,
the canonical or the Weyl-Wigner integral quantization maps $f(q)$ to $f(Q)$
(resp. $f(p)$ to $f(P)$), and so might be unable to cure or regularise a given classical singularity, particularly with regard to the requirement of  essential self-adjointness for basic operators, which is not guaranteed anymore.
This marks one more difference between classical and quantum models. In  Physics one works mostly (if not always!) with effective models, and an effective quantum model is expected, for practical reasons,  to be more regular than its classical one. The latter is often viewed as too mathematically  idealised.

\section{Covariant integral quantization of the motion on the  line}
\label{CIQWH}

In this section, we describe our approach to the quantization of the motion on the line. Precisely, we transform a function $f(q,p) \equiv f(\vr)$ into an operator $A_f$ in some Hilbert space $\mathcal{H}$ through a linear map which sends the function $f=1$ to the identity operator in $\mathcal{H}$ and which respects the basic translational symmetry of the phase space. The trick is to use ressources of measure/integral calculus, where we can ignore points or lines to some extent. A probabilistic content will be one of the most appealing outcomes of the procedure.  

\subsection{The quantization map}
 
We define the integral quantization of the motion on the  line  as  the linear map
\begin{equation}
\label{qmap}
f(\vr) \mapsto A_f = \int_{\R^2} f(\vr)\, \mathfrak{Q}(\vr)\, \frac{\ud^2\vr}{2\pi c_{\mathfrak{Q}}}\, ,\quad \ud^2\vr=\ud q\,\ud p\, . 
\end{equation}
where $c_{\mathfrak{Q}}$ is a positive constant, whose  meaning will be given later, and 
$\mathfrak{Q}(\vr)/(2\pi c_{\mathfrak{Q}})$ is a family of operators   which solve the identity in $\mathcal{H}$ with respect to the Lebesgue measure $\ud^2\vr$,
\begin{equation}
\label{resid}
\int_{\R^2}  \mathfrak{Q}(\vr)\, \frac{\ud^2\vr}{2\pi c_{\mathfrak{Q}}} = I\,. 
\end{equation}
Hence  the identity  $I$  is the quantized version of the function $f=1$. 
It is clear that we can ignore the  immediate solution    $\mathfrak{Q}(\vr)= w(\vr) I$ with $$\int_{\R^2}  w(\vr)\, \dfrac{\ud^2\vr}{2\pi c_{\mathfrak{Q}}}=1\, , $$ which  leads to the trivial quantization  
 \begin{equation*}
A_f = \lg f\rg_{w}\,I\quad \mbox{with} \quad  \lg f\rg_{w}:=\int_{\R^2} f(\vr) \, w(\vr)\, \frac{\ud^2\vr}{2\pi c_{\mathfrak{Q}}}\,,
\end{equation*}
i.e., to the classical statistical mechanics where $w(\vr)$ is  chosen as  a distribution function with respect to the measure $\ud^2\vr/(2\pi c_{\mathfrak{Q}})$.

In addition to \eqref{resid}, we impose  the family $ \mathfrak{Q}(\vr)$ to obey a symmetry condition issued from the homogeneity of the phase space. 
Indeed, the choice of the origin in $\R^2$ is arbitrary. Hence we must have translational covariance in the sense that the quantization of the translated of $f$ is unitarily equivalent to the quantization of $f$ 
\begin{equation}
\label{covtrans1}
U(\vr_0)\,A_f \,U(\vr_0)^{\dag}= A_{\mathcal{T}(\vr_0)f}\, , \quad \left(\mathcal{T}(\vr_0)f\right)(\vr):= f\left(\vr-\vr_0\right) 
\end{equation} }
 So $\vr\mapsto U(\vr)$ has to be a unitary, possibly projective,  \underline{representation} of the abelian group $\R^2$

Then, from \eqref{covtrans1} and the translational invariance of $\ud^2\vr=\ud q\,\ud p$, the operator valued function  $\mathfrak{Q}(\vr)$ has to obey 
\begin{equation}
\label{covQ}
U(\vr_0)\,\mathfrak{Q}(\vr) \,U^{\dag}(\vr_0)= \mathfrak{Q}\left(\vr + \vr_0\right) 
\end{equation}
A solution to \eqref{covQ} is  found by picking an operator $\mathfrak{Q}_0\equiv \mathfrak{Q}(\pmb{0})$ and write 
\begin{equation}
\label{solQ0}
\mathfrak{Q}\left(\vr\right) := U(\vr)\,\mathfrak{Q}_0 \,U^{\dag}(\vr) 
\end{equation}
Then the resolution of the identity holds from Schur's Lemma  \cite{barracz77} if $U$ is irreducible, and if the operator-valued integral  \eqref{resid} makes sense, i.e., if the choice of the fixed operator  $\mathfrak{Q}_0$ is valid.

 \subsection{Toward projective unitary irreducible representations of $\R^2$}
 
 Any  unitary representation $\vr\mapsto U(\vr)$ of the abelian group $\R^2$ has the following properties
\begin{equation}
\label{URR2}
U(\pmb{0}) = I\, , \quad U^{\dag}(\vr) = U(-\vr)\, , \quad U(\vr)\,U(\vrp)= U(\vr + \vrp)\, . 
\end{equation}
Therefore, any true unitary irreducible representation of $\R^2$ is one-dimensional (Fourier!):
\begin{equation}
\label{UIRR2}
\vr \mapsto U_{\pmb{k}}(\vr) = e^{\ii \pmb{k}\cdot\vr}= e^{\ii(k_1q + k_2p)}\, . 
\end{equation}
Now, if we pick one of these representations,  then the  integral quantization that  it defines from \eqref{solQ0} is barred due to absence of resolution  of the identity, 
\begin{equation}
\mathfrak{Q}(\vr) = \mathfrak{Q}_0 \, , \quad \Rightarrow\quad  \int_{\R^2}  \mathfrak{Q}(\vr)\, \frac{\ud^2\vr}{2\pi} = \infty\, . 
\end{equation}
The alternative is to deal with \textbf{projective} unitary representation of $\R^2$ of the form
\begin{align}
\label{prURR2}
U(\pmb{0}) &= I\, , \quad U^{\dag}(\vr) = U(-\vr)\, , \\
 U(\vr)\,U(\vrp)& = e^{\ii \xi(\vr,\vrp)}U(\vr+\vrp)
\end{align}
where the real valued  $\xi$ is accountable for the  non commutativity of the representation, a central feature of the family of  the $A_f$'s, 
\begin{equation}
\label{nsymxi}
\xi(\vr,\vrp) \neq \xi(\vrp,\vr)\, . 
\end{equation}
This function has to fulfil  cocycle conditions which agree with the group structure of $\R^2$ defined by  the relations
\begin{align}
\label{pgstruct1}
U(\pmb{0}) &= I\, , \quad U^{\dag}(\vr) = U(-\vr)\, , \\
\label{pgstruct2} U(\vr)\,U(\vrp)& = e^{\ii \xi(\vr,\vrp)}U(\vr+\vrp)\, ,  
\end{align}
and which determine the function $\xi$.  
We deduce from neutral element and inverse in \eqref{pgstruct1}
\begin{equation}
\label{propxi1}
\xi(\vr,\vz)= 0 = \xi(\vz,\vr)\, , \quad \xi(\vr,-\vr)= 0 = \xi(-\vr,\vr)\, . 
\end{equation}
From associativity $U(\vr)\,(U(\vrp)\,U(\vrpp))= (U(\vr)\,U(\vrp))U(\vrpp)$ we have
\begin{equation}
\label{propxi2}
\xi(\vr,\vrp) + \xi(\vr + \vrp,\vrpp)  = \xi(\vr,\vrp + \vrpp)  + \xi(\vrp,\vrpp)\,.  
\end{equation}
From the Lie group structure of $\R^2$, the function  $\xi$ has to be smooth.  Let us apply  $\left.\nabla_{\vrpp}\right\vert_{\vrpp=\vz}$ to \eqref{propxi2}, and  
 define $\nabla_{\vb}\xi(\va,\vb):= \pmb{F}(\va,\vb)$. We then obtain the functional equation for $\pmb{F}$:
\begin{equation}
\label{FFF}
 \pmb{F}(\vr,\vrp) =  \pmb{F}(\vr+\vrp,\vz)- \pmb{F}(\vrp,\vz)\, , 
\end{equation}
whose solution is the linear $\pmb{F}(\vr,\vrp) = k \vr, $ for some constant $k$. 
It follows that $\xi(\vr,\vrp) $ is  bilinear in $(\vr,\vrp)$. From $\xi(\vr,-\vr) = - \xi(\vr,\vr)= 0$, the only possibility is that $\xi(\vr,\vrp) $ is the symplectic form
\begin{equation}
\label{detxi}
\xi(\vr,\vrp) = k\,(qp^{\prime}-q^{\prime}p)\equiv k\,\vr\wedge\vrp\, . 
\end{equation}
Keeping physical dimensions,  the constant $k$ should read $k= 1/\ell\wp  \neq 0$, where $\ell$ (resp. $\wp$) is some characteristic  length (resp. momentum) appropriate to the scale of the model. Thinking of quantum  systems, we naturally  introduce  the Planck constant $\hbar$ such that $\ell\wp = \hbar$.

\subsection{From $\R^2$ to the Weyl-Heisenberg group and its UIR}
  \label{R2WH} 
Because of the non triviality of $\xi$, we have now to deal with the Weyl-Heisenberg (WH) group,
\begin{equation}
\label{WHgroup}
\mathrm{WH} = \{(s,\vr)\, , \, s\in \R\, , \, \vr\in \R^2\}\, , \quad (s,\vr)(s^{\prime},\vrp)= \left(s+s^{\prime} + \frac{1}{2}\xi(\vr,\vrp), \vr+\vrp\right)\, ,
\end{equation}
instead of just $\R^2$. 

From von Neumann \cite{vneumann31,perel86,combescure12}, WH has a unique non trivial UIR, up to equivalence corresponding precisely to the arbitrariness in the choice of $k$:
\begin{equation}
\label{UIRWH}
(s,\vr) \mapsto \mathcal{U}(s,\vr)= e^{\ii s}\,U(\vr) = e^{\ii s}\, e^{\ii (pQ-qP)/k} 
\end{equation}
where $Q$ and $P$ are the two above-mentionned self-adjoint operators in $\mathcal{H}$ such that $[Q,P]= \ii\, \hbar I$.
 In the present context, $U(\vr)$ is named \textit{displacement operator}. In the sequel we fix $k=1= \hbar$ for convenience, so that  $U(\vr)= e^{\ii (pQ-qP)}$\,. 

 \subsection{WH covariant integral quantization(s)}

From Schur's Lemma applied to the WH UIR $\mathcal{U}$, or equivalently to $U$ since $e^{\ii s}$ is just a phase factor, we confirm the  resolution of the identity 
\begin{equation}
\label{resunit2}
\int_{\R^2}  \mathfrak{Q}(\vr)\, \frac{\ud^2\vr}{2\pi c_{\mathfrak{Q}_0}} = I\,, \quad \mathfrak{Q}(\vr)= U(\vr)\mathfrak{Q}_0 U^{\dag}(\vr)\, , 
\end{equation}
where $\mathfrak{Q}_0$ is the fixed operator introduced in \eqref{solQ0}, whose choice is left to us, and which is such that   $0<c_{\mathfrak{Q}_0}< \infty$. Let us prove  that this is  possible if $\mathfrak{Q}_0$ is trace class, i.e., $\mathrm{Tr}(\mathfrak{Q}_0)$ is finite. Indeed, let us introduce the function
\begin{equation}
\label{WHtr}
\Pi(\vr) = \mathrm{Tr}\left(U(-\vr)\mathfrak{Q}_0 \right)\, , 
\end{equation}
which can be interpreted as the \textit{Weyl-Heisenberg transform} of operator $\mathfrak{Q}_0$. 

The \textit{inverse WH-transform} exists due to two remarkable properties \cite{bergaz13,becugaro17} of the displacement operator $U(\vr)$,
\begin{equation}
\label{IWHtr}
 \int_{\R^2} U(\vr) \,\frac{\ud^2\vr}{2\pi}= 2{\sf P}\ \mbox{and}\ \mathrm{Tr}\left(U(\vr)\right)= 2\pi \delta(\vr) \ \Rightarrow \
\mathfrak{Q}_0 = \int_{\R^2} U(\vr) \, \Pi(\vr)\,\frac{\ud^2\vr}{2\pi}\, , 
\end{equation}
where ${\sf P}= {\sf P}^{-1}$ is the parity operator defined as ${\sf P}U(\vr){\sf P}= U(-\vr)$
 
The function $\Pi(\vr)$ is like a weight, not necessarily normalisable, or even positive. It can be viewed as an apodization \cite{apodi}  on the plane, which determines the extent of our coarse graining  of the phase space.  In a certain sense this  function  corresponds to the Cohen ``$f$'' function \cite{cohen66} (for more details see \cite{cohen13} and references therein) or to Agarwal-Wolf filter functions \cite{agawo70}, even though  these authors were not directly concerned with quantization procedures. 
 
The value of constant $c_{\mathfrak{Q}_0}$ derives from the above and reads
\begin{equation}
\label{cQ0 }
c_{\mathfrak{Q}_0} = \mathrm{Tr}\left(\mathfrak{Q}_0 \right) = \Pi\left(\vz\right)\,.
\end{equation}

Equipped with one choice of a traceclass $\mathfrak{Q}_0$,  we can now proceed with the corresponding WH covariant integral quantization map
\begin{equation}
\label{fAf}
 f(\vr) \mapsto A_f \equiv A^{\mathfrak{Q}_0}_f= \int_{\R^2} f(\vr) \mathfrak{Q}(\vr)\, \frac{\ud^2\vr}{2\pi c_{\mathfrak{Q}_0}}\, . 
\end{equation}
In this context, the operator $\mathfrak{Q}_0$ is the quantum version (up to a constant) of the origin of the phase space, identified with the $2\pi\times$ Dirac distribution at the origin. 
\begin{equation}
\label{quantdirac}
2\pi \delta(\vr) \mapsto A_{\delta}= \frac{\mathfrak{Q}_0}{c_{\mathfrak{Q}_0}}\, , \quad 2\pi \delta(\vr-\vr_0) \mapsto A_{\delta_{\vr_0}}= \frac{\mathfrak{Q}(\vr_0)}{c_{\mathfrak{Q}_0}}
\end{equation}

 \section{Permanent issues of WH covariant integral quantizations}  
 \label{permissu}
 
 By permanent issues we mean that some basic rules managing the quantum model have a kind of universality, almost whatever the choice of admissible  $\mathfrak{Q}_0$, or its corresponding  apodization $\Pi(\vr)$.
 
 \subsubsection*{Symmetric operators and self-adjointness}
First, we have  the general important outcome:  if $\mathfrak{Q}_0$ is symmetric, i.e. $\overline{\Pi(-\vr)}= \Pi(\vr)$,  a real function $f(\vr)$ is mapped to a symmetric operator $A_f$. Moreover, if  $\mathfrak{Q}_0$ is a positive operator, then  a real semi-bounded  function $f(\vr)$ is mapped to a self-adjoint operator $A_f$ through the Friedrich extension \cite{akhglaz81} of its associated semi-bounded quadratic form.   

\subsubsection*{Position and Momentum}
Canonical commutation rule is preserved:
\begin{equation}
\label{qandp}
A_q = Q + c_0\, , \quad A_p= P+d_0\,, \quad c_0,d_0\in \R\, ,  \Rightarrow \left[A_q,A_p\right]= \ii I\, . 
\end{equation}
This result is actually the direct consequence of the underlying Weyl-Heisenberg covariance when one expresses Eq.\eqref{covtrans1} on the level of infinitesimal generators.

 \subsubsection*{Kinetic energy}
\begin{equation}
\label{p2}
 A_{p^2}= P^2 + e_1\,P + e_0\, , \quad e_0, e_1 \in \R
\end{equation}
\subsubsection*{Dilation} 
\begin{equation}
\label{qp}
A_{qp} = A_q\,A_p + \ii f_0\, , \quad  f_0\in \R
\end{equation}
In the above formulas,  the constants $c_0,d_0,e_0,e_1,$ can be easily removed by imposing mild constraints on $\Pi(\vr)$. Moreover, constant $f_0$ can be fixed to $-1/2$ in order to get the symmetric dilation operator $(QP + PQ)/2$. 
\subsubsection*{Potential energy} A potential energy $V(q)$ becomes a multiplication operator in position representation.
\begin{equation}
\label{Vq}
A_{V} = \mathfrak{V}(Q)\, , \quad \mathfrak{V}(Q)= \frac{1}{\sqrt{2\pi}}\,V\ast \overline{\mathcal{F}}[\Pi(0,\cdot)](Q)\,
\end{equation}
where $ \overline{\mathcal{F}}$ is the inverse 1-D Fourier transform, and ``$\ast$'' stands for convolution with respect to the second variable. The case of  singular potentials, e.g., $V(q) = 1/\vert q\vert$, might request  support conditions on  $\overline{\mathcal{F}}[\Pi(0,\cdot)]$. 

\subsubsection*{Functions of $p$}
If $F(\vr)\equiv h(p)$ is a function of $p$ only, then $A_h$ depends on $P$ only through the convolution:
\begin{equation}
\label{hp}
A_h= \frac{1}{\sqrt{2\pi}}\,h\ast \overline{\mathcal{F}}[\Pi(\cdot,0)](P)\, .
\end{equation}

\section{Most reasonable Hamiltonians in Galilean Physics}
\label{mvarhamil}
 For the motion of an interacting massive  particle on the line, it is reasonable to impose the validity of the so-called \textit{shadow} Galilean invariance \cite{jmll74}, \cite{jmll92,jmll95}), which is a nice way to understand gauge invariance: no discrimination is possible instantaneously between a free and an interacting system.  Let us give an account of the reasoning. In the classical context, the phase space $\R^2= \{(q,p)\}$ is an homogeneous space for the  $1+1$ Galileo group $G$ and its extended version $\widetilde{G}$ \cite{jmll74}. We recall that a general active Galilean transformation $(b,a,v)$ of space-time events $(x,t)$ is defined by
 \begin{equation}
\label{stgaltr}
\begin{split}
x &\mapsto x +vt +a\,,\\
 t &\mapsto t +b \, ,
\end{split}
\end{equation}
with the composition law $(b,a,v)(b^{\prime},a^{\prime},v^{\prime}) = (b+b^{\prime}, a+a^{\prime} + vb^{\prime},v+v^{\prime})$, and inverse $(b,a,v)^{-1}= (-b,-a+bv,-v)$. The latter defines the passive transformation $(x,t)\mapsto (x -v(t-b)-a,t-b)$. 
The corresponding infinitesimal generators read respectively, $\mathcal{t}$ for time translations ($t \mapsto t +b$), $\mathcal{p}$ for space translations ($x \mapsto x + a$), and $\mathcal{k}$ for instantaneous Galilean transformations or boosts ($x \mapsto x +vt$). They form  the  Galileo Lie algebra
\begin{equation}
\label{Galie}
[\mathcal{t},\mathcal{p}] = 0\, , \quad  [\mathcal{k},\mathcal{p}]= 0\, , \quad  [\mathcal{k},\mathcal{t}] = \mathcal{p}\, . 
\end{equation}
However, a consistent Galilean description of the motion of a particle of mass $m>0$ necessitates to \textit{centrally extend} the Galilean transformations with the adding of an extra parameter, like we did in Subsection \ref{R2WH} where we  extended the abelian $\R^2$ to the Weyl-Heisenberg group. The extended Galileo group becomes the set of four-parameter elements $(\vartheta,  b,a,v)$, with the composition law 
\begin{equation}
\label{extGlaw}
(\vartheta, b,a,v)(\vartheta^{\prime},b^{\prime},a^{\prime},v^{\prime}) = \left(\vartheta + \vartheta^{\prime} + m va^{\prime} + \frac{1}{2}mv^2b^{\prime},b+b^{\prime}, a+a^{\prime} + vb^{\prime},v+v^{\prime}\right)\, . 
\end{equation}
 We must now add to the three above generators the identity $\mathcal{i}_d$, which corresponds to the phase $\vartheta$, and which commutes with all generators. The extended commutation rules read
\begin{equation}
\label{Galie}
[\mathcal{t},\mathcal{p}] = 0\, , \quad  [\mathcal{k},\mathcal{p}]= m\mathcal{i}_d\, , \quad  [\mathcal{k},\mathcal{t}] = \mathcal{p}\, . 
\end{equation}
While the space-time can be  identified with the group coset $\widetilde{G}/\Theta \times V \sim \R^2$, where the subgroup $\Theta \times V $ consists of phase changes and boosts,  the phase space for the motion of the particle is naturally identified with the coset $\Gamma= \widetilde{G}/\Theta \times T \sim \R^2$ where $T$ is the subgroup of time translations. From the factorization 
\begin{equation}
\label{factorextG}
(\vartheta, b,a,v)= (0, 0,a -vb,v)\,\left(\vartheta - \frac{1}{2}mv^2b, b,0,0\right)\, 
\end{equation}
this coset can be given the  global coordinates $(q,p):= (a-bv, mv)$. It is acted upon by elements $(\vartheta,b^{\prime},a^{\prime},v^{\prime})$ in $\widetilde{G}$ through left multiplication on \eqref{factorextG}. This leads to the transformations:
\begin{equation}
\label{psgaltr}
\begin{split}
q &\mapsto q +a^{\prime} -b^{\prime}v^{\prime} -b^{\prime}\frac{p}{m}\, , \\
p &\mapsto p + m v^{\prime}\, .  
\end{split}
\end{equation}
In this phase space context, the four generators of $\widetilde{G}$ are represented by  the  basic  functions  $g_i(q,p)$, $i=1,2,3,4$, 
\begin{equation}
\label{genflow}
\mathcal{i}_d\mapsto g_1(q,p)=1\, , \quad \mathcal{t}\mapsto \sfH(q,p)\, , \quad \mathcal{p}\mapsto p\,, \quad \mathcal{k}\mapsto K(q,p)\, .
\end{equation}
They  generate the corresponding Galilean flows through Poisson brackets,
\begin{equation}
\label{poissonbr}
\frac{\ud f}{\ud \lambda_i}  = \{f,g_i\}:= \frac{\partial f}{\partial q}\,\frac{\partial g_i}{\partial p}-\frac{\partial f}{\partial p}\,\frac{\partial g_i}{\partial q}\, .
\end{equation}
They  realize the extended Galileo Poisson-Lie algebra, consistently to \eqref{Galie}, in the case of the free particle, 
\begin{equation}
\label{PoissLie}
\{p,\sfH\} = 0\, , \quad  \{K,p\}= m \, , \quad  \{K,\sfH\} = p\, ,
\end{equation}
whose solutions for $\sfH$ and $K$ read $\sfH= \dfrac{p^2}{2m} + U$. Here, the constant $U$ may be viewed as an internal energy,  and $K= mq + \phi(p)$. Now,  the boost is expected  not to modify the position at the time it is performed, and so   
\begin{equation}
\label{boostq}
\phi(p) =0 \Rightarrow \{K,q\} = 0\, .
\end{equation}
One notices from these results that the observable velocity, defined as $V= \dfrac{\ud q}{\ud t} = \{q,\sfH\}= \dfrac{p}{m}$, obeys the canonical commutation rule,
\begin{equation}
\label{VK}
\{K,V\} = 1\, . 
\end{equation}
This means that the boost flow acts on the  velocity as a translation. This formula is the key for getting the expression of  the boost $K$ and the Hamiltonian $\sfH$ when the particle is no longer free. Following Levy-Leblond, we understand  that, even  in presence of interaction, instantaneous Galilean transformations change the velocity without modifying the position.  Hence, \eqref{boostq} and \eqref{VK} remain true. The first one implies that $K$ is a function of $q$ alone, $K=N(q)$, and the second one allows to determine the form of the Hamiltonian $\sfH = \sfH(q,p)$, since we should have $1= \{N,V\} = \{N, \{q,\sfH\}\} $. From the Jacobi identity we have:
\begin{equation}
\label{KVH}
0= \{N, \{q,\sfH\}\} +  \{\sfH, \{N,q\}\} + \{q, \{\sfH,N\}\}= 1 + 0  -  N^{\prime}(q)\,\partial^2_p \sfH\, , 
\end{equation}
This  leads to the expression
\begin{equation}
\label{HMV}
\sfH= \frac{p^2}{2N^{\prime}(q)} + R(q)p + S(q)\, . 
\end{equation}
Thus, we can interpret $N^{\prime}(q)$ as a variable mass $N^{\prime}(q)\equiv m(q)$, and this interpretation is consistent with the commutator $\{K,p\} = m(q)$, which becomes  $\{K,p\} = m=$cst in the non-interacting particle case. 
 
One can conclude, after introducing the evolution parameter $t$,  that shadow Galilean   dynamics is  ruled by  Hamiltonians of the general form,  
\begin{equation}
\label{hamiltGal1}
\begin{split}
\sfH^{\mathrm{gen}} = \sfH^{\mathrm{gen}}(q,p;t) &= \frac{1}{2m(q)}(p - A(q;t))^2 + U(q;t)\\
&=  \frac{p^2}{2m(q)} -\frac{p}{m(q)}\,A(q;t) + A^2(q;t) + U(q;t) \\ &\equiv L_2(q;t)\,p^2 + L_1(q;t)\,p + L_0(q;t)\, , 
\end{split}
\end{equation}
on which our method of integral quantization applies easily, and plays in general a regularizing r\^ole, depending on the choice of the weight $\Pi(\vr)$. Note that this choice will dispel the ordering ambiguity due to the presence of the variable mass.

 \section{Examples of weight functions}
 \label{weight}

 The simplest choice is   $\Pi(\vr) = 1$, of course.  Then $\mathfrak{Q}_0 = 2 \sfP$ and $c_{\mathfrak{Q}_0}= 1$. This no filtering choice yields the popular Weyl-Wigner integral quantization (see \cite{combescure12} and references therein), equivalent to the standard ($\sim$ canonical) quantization. No  regularisation of space or momentum singularity present in the classical model is possible since 
\begin{equation}
\label{VqVQ}
V(q) \mapsto A_V = V(Q)\, , \quad h(p) \mapsto A_h = h(P)\, . 
\end{equation}
This quantization yields the so-called Weyl  ordering \cite{becugaro17}.
Another, less popular choice, is the Born-Jordan weight,  $\Pi(q,p) = \dfrac{\sin qp}{qp}$, which presents appealing aspects \cite{cordero_etal15}. Nevertheless, with this choice Eqs. \eqref{VqVQ}  still hold true.

An easily manageable choice concerns  separable weight $\Pi(q,p)= \lambda(q)\, \mu(p)$, where $\lambda$ and $\mu$ are preferably regular, e.g., rapidly decreasing smooth functions. Such an option is  suitable for physical Hamiltonians which are sums of terms like $L(q)\,p^m$, where it  allows  regularisations through convolutions if functions $\lambda$ and $\mu$ are regular enough. 
\begin{equation}
\label{sepweight}
A_{L(q)\,p^n}=  \sum_{\substack{
         r,s,t\\
         r+s+t=n}}  2^{-s}\, \binom{n}{r\,s\,t}\,\ii^r\,\lambda^{(r)}(0)\,(-\ii)^s \frac{1}{\sqrt{2\pi}}\,\left(\overline{\mfF}[\mu]\ast L \right)^{(s)}(Q)\, P^t\, . 
\end{equation} 
Note that $\left(\overline{\mfF}[\mu]\ast L \right)^{(s)}= \left(\overline{\mfF}[\mu]\right)^{(s)}\ast L = \overline{\mfF}[\mu]\ast L^{(s)}$, relations whose  validity depends on the derivability of the factors.   

For the cases $n=0$, $n=1$ and $n=2$, i.e. the most relevant to  Galilean physics, we have, with $T(x):= \frac{1}{\sqrt{2\pi}}\,\left(\overline{\mfF}[\mu]\ast L \right)(x)$, 
\begin{equation}
\label{Lpm0}
A_{L(q)} = \lambda(0)\, T(Q)
\, ,
\end{equation}
\begin{equation}
\label{Lpm1}
\begin{split}
A_{L(q)\,p} &= \lambda(0)\, T(Q)\, P + \ii \,\lambda^{\prime}(0)\, T(Q) -\frac{\ii}{2}\,\lambda(0)\, T^{\prime}(Q)\\
&= \lambda(0)\, \frac{T(Q)\, P + P\,T(Q)}{2} + \ii\, \lambda^{\prime}(0)\, T(Q)\, ,
\end{split}
\end{equation}
\begin{align}
\label{Lpm2a} A_{L(q)\,p^2} &= \lambda(0)\, T(Q)\, P^2 + \ii \,(2\lambda^{\prime}(0)\, T(Q)-  \lambda(0)\, T^{\prime}(Q))\,P \\ 
 \nonumber &-\lambda^{\prime\prime}(0)\, T(Q)+ \lambda^{\prime}(0)\, T^{\prime}(Q)-\frac{\lambda(0)}{2}\, T^{\prime\prime}(Q)\\
 \nonumber&= \lambda(0)\, \frac{T(Q)\, P^2 + P^2\,T(Q)}{2} + 2\ii\, \lambda^{\prime}(0)\, T(Q)\,P\\ \nonumber &-\lambda^{\prime\prime}(0)\, T(Q)+ \lambda^{\prime}(0)\, T^{\prime}(Q) + \frac{\lambda(0)}{4}\, T^{\prime\prime}(Q)\, .
\end{align} 
We observe that the  operators \eqref{Lpm1} and \eqref{Lpm2a}  are symmetric under the condition 
\begin{equation}
\label{la0}
\lambda^{\prime}(0) = 0\, . 
\end{equation} 
Note  the appearance, in the expression of the operator \eqref{Lpm2a}, of a potential built from derivatives of the regularisation of $L(q)$. This feature is typical of quantum Hamiltonians with variable mass (see the discussion in \cite{jmll95}).

Separable Gaussian weights $\Pi(q,p) = e^{-\frac{q^2}{2\sigma_{\ell}^2}}\, e^{-\frac{p^2}{2\sigma_{\eth}^2}}$
 yield simple formulae with familiar probabilistic content. Moreover they satisfy Condition \eqref{la0}. Standard coherent state (or Berezin or anti-Wick)  quantization corresponds to the particular values $\sigma_{\ell}=\sqrt{2}=\sigma_{\eth}$. 
The limit Weyl-Wigner case holds as  the widths $\sigma_{\ell}$ and $\sigma_{\eth}$ are infinite (Weyl-Wigner is singular in this respect).

\section{Probabilistic content}
\label{probcont}
 The probabilistic content of our quantization procedure is better captured if one uses an  alternative quantization formula through the so-called symplectic Fourier transform. The latter is defined as 
  \begin{equation}
\label{symFourqp}
 \mathfrak{F_s}[f](\vr)= \int_{\R^2}e^{-\ii \vr\wedge\vrp}\, f(\vrp)\,\frac{\ud^2\vrp}{2\pi} \, . 
\end{equation}
 It is involutive, $\mathfrak{F_s}\left[\mathfrak{F_s}[f]\right]=  f$ like its \textit{dual} defined as $\overline{\mathfrak{F_s}}[f](\vr)= \mathfrak{F_s}[f](-\vr)$. 
 
The equivalent form of  the WH integral quantization \eqref{fAf} reads as
\begin{equation}
\label{quantPi1}
f\mapsto A_f= \int_{\R^2}  U(\vr)\,  \overline{\mFs}[f](\vr)\, \frac{\Pi(\vr)}{\Pi\left(\vz\right)} \,\frac{\ud^2 \vr}{2\pi}\, .  
\end{equation}
This formula allows to prove  an interesting trace formula (when applicable to $f$): 
\begin{equation}
\label{traceAf}
\mathrm{Tr}\left(U(\vr)\right)=  2\pi \delta(\vr) \Rightarrow \mathrm{Tr}\left(A_f\right)=  \overline{\mFs}[f](\vz)=  \int_{\R^2}    f(\vr)\,  \,\frac{\ud^2 \vr}{2\pi}\, .
\end{equation}

By using \eqref{traceAf} we derive the 
\textit{quantum phase space portrait} of the operator as an autocorrelation averaging of the original $f$. 
 More precisely, starting from a function (or distribution) $f(\vr)$, one defines through its quantum version $A_f$ the new function $\check f(\vr)$ as
\begin{equation}
\label{fmapcf}
\check f(\vr) =\frac{1}{c_{\mathfrak{Q}_0}} \mathrm{Tr}\left(\mathfrak{Q}(\vr)A_f\right)=\int_{\R^2}  \, \frac{\mathrm{Tr}\left(\mathfrak{Q}(\vr)\,\mathfrak{Q}(\vrp)\right)}{c^2_{\mathfrak{Q}_0}}\, f(\vrp)\frac{\ud^2 \vrp}{2\pi}\, .  
\end{equation}
The map $\vrp\mapsto  \dfrac{\mathrm{Tr}\left(\mathfrak{Q}(\vr)\,\mathfrak{Q}(\vrp)\right)}{c^2_{\mathfrak{Q}_0}}$ might be a probability distribution if this expression is non negative.  Now, this map is better understood from the equivalent formulas,
\begin{equation}
\label{fmapcf1}
\begin{split}
\check f(\vr)  &= \int_{\R^2}  \mFs\left[\frac{\Pi\,\widetilde\Pi}{\Pi^2(\vz)}\right](\vrp-\vr)\, f(\vrp) \,\frac{\ud^2 \vrp}{2\pi} \\
&=\int_{\R^2}  \mFs\left[\frac{\Pi}{\Pi\left(\vz\right)}\right]\ast\mFs\left[\frac{\widetilde\Pi}{\Pi\left(\vz\right)}\right](\vrp-\vr)\, f(\vrp) \,\frac{\ud^2 \vrp}{4\pi^2} 
\end{split}
\end{equation}
 This represents the convolution ($\sim$ local averaging)  of the original $f$ with the autocorrelation of the symplectic Fourier transform of the (normalised) weight $\dfrac{\Pi(\vr)}{\Pi\left(\vz\right)}$. 
 
Hence, in view of the above convolution, we are incline to choose windows $\Pi(\vr)$, or equivalently $\mathfrak{Q}_0$,  such that
\begin{equation}
\label{probdist}
 \mFs\left[\frac{\Pi}{\Pi\left(\vz\right)}\right]
\end{equation}
is a probability distribution on the plane $\R^2$ equipped with the measure $\dfrac{\ud^2 \vr}{2\pi}$. 
A sufficient condition is that  $\mathfrak{Q}_0$ is a density operator, i.e., non-negative and unit trace. 
It is not necessary, since 
the uniform Weyl-Wigner choice  $\Pi(\vr)= 1$ yields 
\begin{equation}
\label{WWdist}
 \mFs\left[1\right](\vr)= 2\pi \delta(\vr)
\end{equation}
and $\mathfrak{Q}_0 = 2\mathrm{P}$, which is not a density operator. Note that $\check f= f$ in this case. Also note  that  the celebrated Wigner function  $\mathcal{W}_{\rho}(\vr)$ for a density operator or mixed quantum state $\rho$, defined by 
\begin{equation}
\label{wigA}
\mathcal{W}_{A}(\vr) = \mathrm{tr}\left(U(\vr)2{\sf P}U^{\dag}(\vr)A\right)\, ,
\end{equation}
is a normalised quasi-distribution which can assume negative values.

With a true probabilistic content,  the meaning of the convolution
\begin{equation}
\label{truedist}
 \mFs\left[\frac{\Pi}{\Pi\left(\vz\right)}\right]\ast\mFs\left[\frac{\widetilde\Pi}{\Pi\left(\vz\right)}\right]
\end{equation}
is clear: it is the probability distribution  for the difference of two vectors in the phase plane, viewed as independent random variables,  and thus is perfectly adapted to the abelian and homogeneous structure of the classical phase space.

 We can conclude that a quantum phase space portrait in this probabilistic context  is like a measurement of  the intensity of a  diffraction pattern resulting from the $\Pi$ coarse graining of the idealistic phase space $\R^2$. 

\section{An example of regularisation}
\label{exreg}

As an elementary example, let us pick the one-dimensional model of a particle with position-dependent mass. This model  was considered by Levy-Leblond in \cite{jmll92}.  The motion of the particle is constrained by a  potential barrier $V(q)$, such that the mass $m = m(q)$ also changes at the potential discontinuity, that is:
\begin{equation}
\label{model1}
V(q)=\left\lbrace\begin{array}{cc}
    0  & (q< 0 )  \\
     V_0 &  (q>0) 
\end{array}\right.  \, , \quad m(q)=\left\lbrace\begin{array}{cc}
    m_l  & (q< 0)   \\
     m_r &  (q>0) 
\end{array}\right. \, . 
\end{equation}

We choose the separable Gaussian weight mentioned above,
\begin{equation}
\label{sepgauss}
\Pi(q,p) = e^{-\frac{q^2}{2\sigma_{\ell}^2}}\, e^{-\frac{p^2}{2\sigma_{\eth}^2}}\, , 
\end{equation}
with  arbitrary widths $\sigma_{\ell}$ and $\sigma_{\eth}$. The application of the formulae \eqref{Lpm0} and  \eqref{Lpm2a} yields the quantum version of the Hamiltonian $H(q,p)$ (with $\hbar= 1$),
\begin{equation}
\label{AHstep}
A_H=   \frac{T(Q)\, P^2 + P^2\,T(Q)}{2} + \mathcal{V}_{+}(Q)= P\,T(Q)\, P  + \mathcal{V}_{-}(Q)\, , 
\end{equation}
where
\begin{equation}
\label{TQ}
T(x)= \frac{1}{4}\left(\frac{1}{m_r}- \frac{1}{m_l}\right)\, \mathrm{Erfc}\left(-\frac{\sigma_{\eth}}{\sqrt{2}}\,x\right) + \frac{1}{2m_l}
\end{equation}
and
\begin{equation}
\label{VQ}
\mathcal{V}_{\pm}(x)= \frac{1}{\sigma_{\ell}^2}\, T(x) \pm \frac{1}{4}\, T^{\prime\prime}(x) + \frac{V_0}{2}\,\mathrm{Erfc}\left(-\frac{\sigma_{\eth}}{\sqrt{2}}\,x\right) 
\end{equation}
with 
\begin{equation}
\label{Tpp}
T^{\prime\prime}(x)=- \frac{\sigma_{\eth}^3}{2\sqrt{2\pi}}\left(\frac{1}{m_r}- \frac{1}{m_l}\right)\,x\, \exp\left(-\frac{\sigma^2_{\eth}}{2}\,x^2\right)\, . 
\end{equation}
The error function $\mathrm{Erfc}$ is defined \cite{abraste72} as
\begin{equation}
\label{Erfc}
\mathrm{Erfc}(x) = \frac{2}{\sqrt{\pi}}\int_x^{\infty} \ud t \, e^{-t^2}= 2- \mathrm{Erfc}(-x)\, , \quad \mathrm{Erfc}(x) = \left\lbrace\begin{array}{cc}
  2 \,,    & x= -\infty   \\
   1 \,,  &  x = 0 \\
   0 \,,  &  x= +\infty
\end{array}\right.\, . 
\end{equation}
Due to these specific values assumed by  the error function, we find for the regularised inverse double mass and the quantum potentials at $x=0\, , \, \pm \infty$, 
\begin{equation}
\label{TxVx}
T(x) = \left\lbrace\begin{array}{cc}
    \frac{1}{2m_l} \, ,  & x= -\infty   \\
   \frac{1}{4}\left(\frac{1}{m_l} +  \frac{1}{m_r}\right) \, , &  x = 0 \\
    \frac{1}{2m_r} \, ,  &  x= +\infty
\end{array}\right.\, ,  \quad \mathcal{V}_{\pm}(x) = \left\lbrace\begin{array}{cc}
   \frac{1}{2m_l\sigma^2_{\ell} } \, ,   & x= -\infty   \\
  \frac{V_0}{2} +  \frac{1}{4\sigma^2_{\ell}}\left(\frac{1}{2m_l} +  \frac{1}{2m_r}\right)\, ,    &  x = 0 \\
   V_0 +   \frac{1}{2m_r\sigma^2_{\ell} }  \, ,   &  x= +\infty
\end{array}\right.\, . 
\end{equation}
From \eqref{AHstep} one can notice the difference $T^{\prime\prime}/2$ resulting from the two types of symmetrisation of the kinetic term, the second one being preferentially picked by Levy-Leblond in \cite{jmll92}. Actually, this type of distinction is not  relevant to our case, since $T^{\prime\prime} \to 0 $ as $\sigma_{\eth}\to \infty$, i.e. at the canonical limit. 

\begin{figure}[htb!]
\begin{center}
\includegraphics[width=5in]{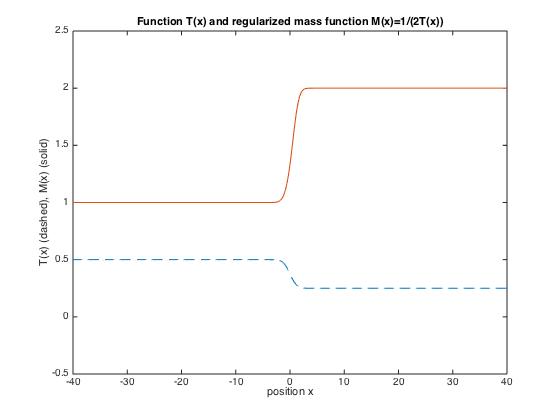}
\includegraphics[width=5in]{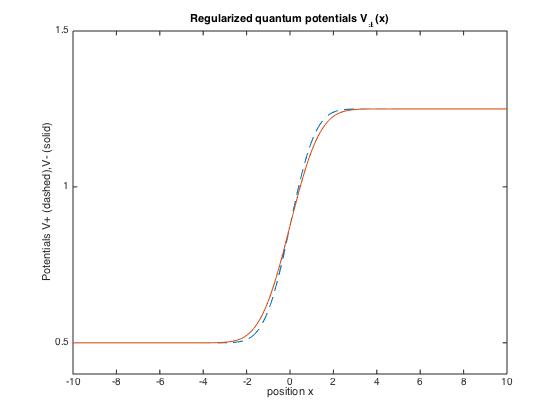}
\caption{On the top: the function $T(x)$ defined by \eqref{TQ} and the corresponding regularized mass $\mathcal{M}(x) = 1/(2T(x))$ replacing  the discontinuous mass introduced in \eqref{model1} with $m_l< m_r$. On the bottom, the regularized potentials $\mathcal{V}_{\pm}(x)$ defined by  \eqref{VQ}.}
\label{masspot}
\end{center}
\end{figure}

In Figure \ref{masspot} are shown  graphs of  the regularised mass $\mathcal{M}(x) = 1/(2T(x))$ and potentials $\mathcal{V}_{\pm}(x)$ for the case $m_l < m_r$. 
Let us now establish the quantum phase space portraits of the Hamiltonian operator $A_H$ along the lines given by \eqref{fmapcf1}.  We obtain the following smooth regularisation of the original $H(q,p)$:
\begin{equation}
\label{semclH}
\check H(q,p)= T(q)\,p^2 + \frac{2}{\sigma_{\ell}^2}\, T(q)  + \frac{V_0}{2}\,\mathrm{Erfc}\left(-\frac{\sigma_{\eth}}{\sqrt{2}}\,q\right)\equiv   T(q)\,p^2 + 
\mathcal{V}_{\mathrm{sc}}(q) \, . 
\end{equation}
We note the factor $2$ appearing in the first term of the semi-classical potential and which is not present in the quantum potential. 
\section{Conclusion}
\label{conclu}

We have outlined a procedure transforming a  classical model for a physical system into one of its quantum versions by using a combination of  symmetry principle, integral calculus on operators and functions, with  a (quasi-) probabilistic interpretation as a guideline. The procedure is applied to  the motion of a variable mass particle on the line, and illustrated with the elementary case of a step potential. The extension to more realistic cases is actually straightforward, save for unescapable technicalities. We would like to promote the idea that in the building of a quantum model, supposed to agree better with observation, one can start  from a  classical rough model or sketch, allowing mathematical idealisations, smoothness, infinities, discontinuities, singularities, and then correct the quantum outcome by using the large freedom we dispose with the choice of a certain coarse-graining determining the procedure. The considered symmetry in the present work was the projective representation of  translations in  the Euclidean  plane, as much rich than it is simple.  Clearly, the method can be adopted in considering many other phase space geometries, e.g., half-plane for the motion on the half-line \cite{gazmur16,albegasca17}, cylinder for the motion on the circle \cite{fregano17}, $\R^2_{\ast}\times \R^2$ for the motion in a punctured plane \cite{gazkoimur17}, etc. 

A promising development of the method \cite{gazkoi17} is to start from $\R^{2n}$ as a phase space, then extend the  method presented in this paper by using the  Weyl-Heisenberg projective translationnal symmetry of $\R^{2n}$, and restrict the quantization to all observables with support in a fixed smooth or singular manifold in $\R^n$.

\subsection*{Acknowledgments}
The author is indebted to  the Centro Brasileiro de Pesquisas F\'{\i}sicas (Rio de Janeiro) and   CNPq Agency (Brazil), and the Institute for Research in Fundamental Sciences (IPM, Tehran) for financial support. He also thanks the  CBPF and the IPM for hospitality. He is grateful to Evaldo M.F.  Curado (CBPF) for valuable comments on the content of this work. 

%
%

\end{document}


\begin{thebibliography}{99}
\bibitem{apodi} An apodization function (also called a tapering function or window function) is a function used to smoothly bring a sampled signal down to zero at the edges of the sampled region.
\end{thebibliography}
\end{document}